\documentclass[12pt]{spieman}  
\usepackage{amsmath,amsfonts,amssymb}
\usepackage{graphicx}
\usepackage{setspace}
\usepackage{tocloft}
\usepackage{lineno}

\usepackage{booktabs}
\usepackage{siunitx}
\usepackage{multirow}
\usepackage{color,soul}

\title{Evaluation of Synthetically Generated CT for use in Transcranial Focused Ultrasound Procedures}

\author[a,$\dagger$]{Han Liu}
\author[b,c,$\dagger$]{Michelle K. Sigona}
\author[b,c]{Thomas J. Manuel}
\author[c,d]{Li Min Chen}
\author[e]{Benoit M. Dawant}
\author[b,c,d,*]{Charles F. Caskey}
\affil[a]{Dept. of Computer Science, Vanderbilt University, Nashville, TN 37235, USA}
\affil[b]{Dept. of Biomedical Engineering, Vanderbilt University, Nashville, TN 37235, USA}
\affil[c]{Institute of Imaging Science, Vanderbilt University, Nashville, TN 37235, USA}
\affil[d]{Dept. of Radiology and Radiological Sciences, Vanderbilt University, Nashville, TN 37235, USA}
\affil[e]{Dept. of Electrical and Computer Engineering, Vanderbilt University, Nashville, TN 37235, USA}
\affil[$\dagger$]{These authors contributed equally to this work}

\cftpagenumbersoff{figure}
\cftpagenumbersoff{table} 
\begin{document} 
\maketitle

\begin{abstract}\\
\textbf{Purpose:} Transcranial focused ultrasound (tFUS) is a therapeutic ultrasound method that focuses sound through the skull to a small region noninvasively and often under MRI guidance. CT imaging is used to estimate the acoustic properties that vary between individual skulls to enable effective focusing during tFUS procedures, exposing patients to potentially harmful radiation. A method to estimate acoustic parameters in the skull without the need for CT would be desirable.\\
\textbf{Approach:}
Here, we synthesized CT images from routinely acquired T1-weighted MRI by using a 3D patch-based conditional generative adversarial network (cGAN) and evaluated the performance of synthesized CT (sCT) images for treatment planning with tFUS. We compared the performance of sCT to real CT (rCT) images for tFUS planning using Kranion and simulations using the acoustic toolbox, k-Wave. Simulations were performed for 3 tFUS scenarios: 1) no aberration correction, 2) correction with phases calculated from Kranion, and 3) phase shifts calculated from time-reversal.\\
\textbf{Results:}
From Kranion, skull density ratio, skull thickness, and number of active elements between rCT and sCT had Pearson’s Correlation Coefficients of 0.94, 0.92, and 0.98, respectively. Among 20 targets, differences in simulated peak pressure between rCT and sCT were largest without phase correction ($12.4\pm8.1\%$) and smallest with Kranion phases ($7.3\pm6.0\%$). The distance between peak focal locations between rCT and sCT was less than 1.3 mm for all simulation cases.\\
\textbf{Conclusions:}
Real and synthetically generated skulls had comparable image similarity, skull measurements, and acoustic simulation metrics. Our work demonstrates the feasibility of replacing real CTs with the MR-synthesized CT for tFUS planning. Source code and a docker image with the trained model are available at \url{https://github.com/han-liu/SynCT_TcMRgFUS}

\end{abstract}


{\noindent \footnotesize\textbf{*}Charles F. Caskey,  \linkable{charles.f.caskey@vanderbilt.edu} }

\begin{spacing}{2}   

\section{Introduction}
\label{sect:intro}  
Transcranial focused ultrasound (tFUS)  is a novel noninvasive method of focusing energy through the skull that often uses Magnetic Resonance Imaging (MRI) for target identification, treatment planning, and closed-loop control of energy deposition\cite{jolesz2009mri}. Focused ultrasound is clinically approved for thermally ablating the thalamus\cite{elias2013pilot} and when used at lower energy levels is being explored for other applications, such as drug delivery and neuromodulation\cite{meng2021applications}. Precise focusing is critical for all tFUS procedures to minimize treatment of off-target tissues\cite{wintermark2014t1}. Prior to tFUS, CT images are acquired to estimate regional skull density, speed of sound, and ultrasound attenuation during ultrasound wave propagation\cite{aubry2003experimental}. Thermally ablative thalamotomy procedures use MR thermometry\cite{https://doi.org/10.1002/jmri.21265} to intraoperatively monitor thermal dose and targeting accuracy. MR Thermometry relies on the temperature dependence of the proton resonance frequency shift to linearly map phase differences between two time points to temperature change. Another tFUS application is neuromodulation, a nonthermal method that has been demonstrated in humans targeting the thalamus\cite{legon2018neuromodulation}, somatosensory cortex\cite{lee2015image}, and primary visual cortex\cite{lee2016transcranial}. During neuromodulation procedures, neuronavigation aids in real-time transducer placement by calculating the position and rotation of optically tracked tools and projecting the transducer's focus onto pre-acquired images. The projected focus from optical tracking is usually a free-field estimate of the focus location\cite{kim2012image}, neglecting the inhomogenous layers of the skull known to shift and distort the focus\cite{sun1998focusing}. The inclusion of CT images to the neuromodulation planning process allows incorporation of skull models to map the skull layers to acoustic properties and estimate spatial accuracy, spatial extent, and output pressure for patient-specific skull models. CT imaging burdens patients by requiring longer screening time and increased risk due to radiation. For tFUS research in development and preclincal phases, it is unrealistic to obtain CT scans of a healthy participant. Therefore, it is desirable to replace the real CT (rCT) images with synthetic CT (sCT) images that are generated from other imaging modalities.

Values from CT images of the head are used in different ways during treatment planning for all tFUS procedures. One important metric is the Skull Density Ratio (SDR), an estimate of the transparency of the skull to ultrasound. The SDR is not always predictive of the energy needed to generate a focal spot transcranially, but a lower SDR is generally interpreted to mean lower acoustic transmission through the skull\cite{boutet2019relevance,chang2016factors}. Although the precise method for computing SDR on a clinical system is proprietary, the metric is derived from the ratio of the Hounsfield Units (HU) of trabecular to cortical bone along the line from a transducer element to the focus\cite{chang2016factors}, and an open source software, Kranion, is available that is capable of generating SDR metrics highly correlated to those found in clinical procedures\cite{sammartino2019kranion}. Along with SDR, Kranion can report a Skull Thickness (ST) measurement between bone layers and Number of Active Elements (NAE), or an element's ray less than $<$ 20 degrees incident to the skull. Detailed spatial maps have been created from CT images to map acoustic properties and model the propagation of sound through the skull \cite{pichardo2010multi,aubry2003experimental,marquet2009non}. Using modeling tools like the acoustic toolbox, k-Wave\cite{treeby2010k}, simulations are used to observe ultrasound waves interacting with subject-specific heterogeneous skulls, quantifying the focal shift, focus size, and energy loss caused by the aberrating skull. 

The use of multi-element arrays during tFUS procedures is desirable because each individual element's amplitude and phase can be precisely controlled. Electronically controlled elements are integral during tFUS procedures to move the transducer's focus location without physically manipulating the transducer and to calculate phase shifts to correct for the skull\cite{kyriakou2014review,meng2021technical}. Several aberration correction methods have been explored that vary in run-time and focus restoration performance\cite{raytrace,histotripsy,sammartino2019kranion}. For clinical thermoablation, 
real-time estimation of amplitude and phase correction are essential as procedures require shifting the small focal volume throughout the brain to ablate the full target\cite{meng2021technical}. Selection of correction method is usually dependent on a trade-off between time constraints and intensity required for a given application.

Deep-learning based methods have been previously used to generate synthetic CTs from MR images\cite{guo2019feasibility}. Dual-echo ultrashort TE (UTE) MR imaging\cite{su2020transcranial} was used to train a 2D U-Net\cite{ronneberger2015u} that was efficient at generating realistic skulls, but UTE scans are not widely available and require development on an MR scanner, as they are not standard protocols. An alternative to UTE images are T1-weighted images, but these can be more challenging to synthesize CT skulls from than UTE because UTE imaging can capture signals from tissues with a very short transverse relaxation time such as bone, providing more information for skull synthesis. For instance, Lei \textit{et al.}\cite{lei2018mri} proposed to use patch-based features extracted from MRIs to train a sequence of alternating random forests based on an iterative refinement model. Maspero \textit{et al.}\cite{maspero2020deep} trained three 2D cGANs\cite{cgan} for each plane and combine the results to generate synthetic CT from T1-weighted MRI. Gupta \textit{et al.}\cite{gupta2019generation} proposed to train a 2D U-Net on sagittal views of MRIs and synthesize the HU of air, soft tissue and bone in three output channels. However, 2D networks are limited by the lack of information of relationship between slices\cite{yu2019thickened} and the skulls are not spatially continuous (i.e generated volumes can appear jagged) along the views that are not involved in training. The irregular skull geometry of sCT may lead to significant differences in tFUS planning. In very recent work done in parallel to ours \cite{koh2021acoustic} a 3D cGAN was proposed to synthesize the whole head CT from MR images. Here, we focus on the skull, which is the critical structure for tFUS.

We hypothesize that synthetic CT generated from MRI can yield comparable clinical metrics for transcranial ultrasound that are derived from CT. Our study used two open-source software tools to compare skull metrics derived from sCT and rCT images using 10 testing cases with two targets. We evaluated the performance of the rCT and sCT skulls using Kranion to report the SDR, ST, and NAEs. Acoustic simulations were performed using k-Wave to calculate the pressure field formed from interactions with each CT and compared the aberration correction performance capabilities through a fast ray-tracing method and a computationally expensive time reversal technique. From each simulation we quantified the maximum intracranial pressure, focal shift between the peak pressure and intended target locations, and focal volumes. Demonstration of similarity between sCT and rCT would show feasibility of synthesizing spatially continuous CT skulls from T1-weighted MRI. 

\section{METHODS}
\subsection{Dataset and Pre-processing}
In this study, our dataset included 86 paired CT and T1-weighted MRI scans of Parkinson patients, who underwent Deep Brain Stimulation from Vanderbilt University Medical Center. Informed consent was obtained from all subjects included in this study. The in-plane resolution of CT images ranged from 0.4297 to 0.5449 mm with a slice thickness of 0.67 mm, while the MR images had an isotropic voxel size of 1 mm. To prepare the paired CT-MRI dataset for network training, we applied a series of image pre-processing procedures as follows. First, for each subject, we spatially aligned the MRI and CT images by rigid registration. Specifically, we registered the low-resolution MRI scans to the high-resolution CT images to preserve the HU values in high-resolution CT images. Rigid registration was used based on the assumption that the shape and size of brain anatomical structures do not vary for the same subject in different imaging modalities. Here, we employed an open-source medical imaging library ANTsPy for rigid registration, where mutual information was used as the cost function. Second, to discard the irrelevant brain regions to our skull synthesis task, we filtered out the non-skull regions from the CT images. Specifically, we extracted a binary mask of the skull region by using an empirically selected threshold, i.e., 400 HU. We then took the largest connected component of the mask to remove other isolated regions. To preserve some contextual information around the skull, we further performed morphological dilation to the skull mask with a ball-shaped structuring element with radius of 4 voxels. This dilated mask was used to filter the raw CT image to obtain the skull-only CT image. Besides, for CT images, we clipped the intensity values to the range of $[-1024, 3071]$ HU and linearly scaled them to $[-1,1]$. For MRI scans, we applied Z-score normalization to each scan, which was further clipped to the $[0.01^{th}, 99.9^{th}]$ percentile of the intensity values, followed by a linearly scaling to $[-1,1]$. 

\subsection{Network Architecture and Training}
As an extension of our previous work\cite{liu2022synthetic}, we adopted a 3D patch-based cGAN to generate a synthetic CT skull image given a T1-weighted MRI. Our 3D cGAN consisted of a generator $G$ and a discriminator $D$, where $G$ was trained to generate realistic CT skulls to fool the discriminator, while $D$ was trained to classify the real and synthetic CT skulls. For the generator $G$, we followed the network architecture in pix2pix\cite{cgan}, i.e., 2D ResNet\cite{he2016deep} with 9 residual blocks, and extended the network to 3D. The residual blocks were useful for shuttling the useful low-level features extracted from MR images, e.g., the location of prominent edges, directly to the deeper layers. At the output layer of $G$, we used Tanh as the activation function to map the logits to a bounded range of $[-1, 1]$. As shown in DCGAN\cite{radford2015unsupervised}, this bounded output range allowed the model to learn more quickly to saturate during the training process. For the discriminator $D$, we adopted a 3D PatchGAN classifier\cite{cgan}, which could be run convolutionally to provide the ultimate output by averaging all responses across the whole volume.

Due to the limit of GPU memory, the network input was a 3D patch $x\in\mathbb{R}^{256\times256\times32}$ randomly cropped from the whole MRI volume $X$. Similarly, the corresponding CT patch $y$ was cropped from the same spatial position from the whole CT volume $Y$. The loss function of cGAN $L_{cGAN}$ could be expressed as 

\begin{equation}
    L_{cGAN}(G,D)=\mathbb{E}_{x,y}[log(D(x,y))]+\mathbb{E}_{x}[log(1-D(x,G(x)))]
\end{equation}

Note that the input of the 3D PatchGAN classifier $D$ was also conditioned on $x$, as it was found critical for GAN to produce realistic outputs \cite{cgan}. A L1 reconstruction loss $L_{L1}$ which encouraged less blurring \cite{cgan} was used and expressed as

\begin{equation}
    L_{L1}(G)=\mathbb{E}_{x,y}[||y-G(x)||_{1}]
\end{equation}

Besides, we included an additional edge-aware loss $L_{edge}$\cite{luo2021edge} to further align the edges between $G(x)$ and $y$. Specifically, we computed the edge maps with a 3D Sobel filter $h(\cdot)$ and minimized the L1 distance between the edge maps. The edge-aware loss function was expressed as

\begin{equation}
    L_{edge}(G)=\mathbb{E}_{x,y}[||h(y)-h(G(x))||_{1}]
\end{equation}

The final objection function $G^{*}$ was the weighted sum of the three loss terms:

\begin{equation}
    G^{*}=\arg \min_{G} \max_{D} L_{cGAN}(G,D) + \lambda_{1}L_{L1}(G) + \lambda_{2}L_{edge}(G)
\end{equation}

We followed the standard training strategy to train our cGAN\cite{goodfellow2014generative}: we updated the parameters of $D$ and $G$ alternatively per gradient descent step. We used the Adam optimizers\cite{kingma2014adam} with an initial learning rate as $2\times 10^{-4}$ and momentum parameters $\beta_{1} = 0.5$ and $\beta_{2} = 0.999$. A mini-batch size of 1 was used. During training, we applied online intensity augmentations\footnote{\url{https://docs.monai.io/en/stable/transforms.html}} including random intensity shifts with an offset ranging from $[-0.1, 0.1]$ and random contrast adjustment with gamma ranging from $[0.5, 1.5]$. Both intensity augmentations were applied with a probability of $0.2$. This augmentation strategy aimed to simulate the varying intensity distributions of MRI scans, especially when they were acquired from different sites or with different protocols.
We did not augment the intensity values of CT images to preserve the physical meaning of the skull HU values. During the inference phase, we used a sliding window to generate synthetic CT patches across the whole volume and fused the results by averaging multiple predictions over each pixel. We set the overlapping ratio between the sliding windows as $0.75$ of the patch dimension, i.e., $192\times192\times24$. For post-processing, we obtained a skull mask from the synthetic CT image following the same steps as in pre-processing, i.e., global threshold, connected component analysis and morphological dilation. This mask was used to remove the false positive predictions outside the skull regions.

\subsection{Study Design}
\subsubsection{Evaluation of Image Similarity}
In our experiment, we randomly split the entire dataset into subgroups containing 66 images for training, 10 for validation and the remaining 10 for testing. We trained the networks by a total number of 1500 epochs and determined the best weighting factors $\lambda_{1}=100$ and $\lambda_{2}=10$ by grid search based on the validation set. To better evaluate the performance of skull synthesis, we only computed the MAE within the skull region of the ground truth. We evaluated the synthesis performance of our 3D patch-based cGAN against the other mainstream MR-CT translation approach, i.e., autoencoder.

\subsubsection{Target Selection}
 The left or right ventral intermediate nucleus (Vim) of the thalamus was used as the target of interest for acoustic evaluation. The right and left Vim segmentation regions were identified by the 'Ventral\_Lateral\_Nucleus' label from the International Consortium Brain Mapping (ICBM) \footnote{\url{http://www.bmap.ucla.edu/portfolio/atlases/ICBM\_Template}} template atlas. The ICBM atlas was registered to each test case MR image with a transformation calculated from an affine registration using 3D Slicer's (version 4.11.2) \footnote{\url{https://www.slicer.org}} General Registration (BRAINS)' module. The output transformation from the registration was applied to the segmented labels and exported as a binary volume as a guide to position the transducer in Kranion.
 
 \subsubsection{Evaluation of Skull Metrics using Kranion}
 In Kranion, we placed a virtual 990-element hemispherical array transducer so that the focus was positioned at the right or left Vim so that there were two targets for each skull in our test dataset of 10 skulls. The transducer geometry is comparable to the the 1024-element ExAblate transducer (Insightec, Tirat Carmel, Israel). The transducer was tilted along the x and y axes and rotated so that the most number of active elements with the sCT skull was displayed on Kranion's graphical interface, without exceeding 10 degrees in each direction to simulate a realistic scenario. The average and cortical bone speeds were maintained as 2705 m/s and 3103 m/s. Skull measurements, transducer element coordinates, and the focus coordinates were exported after positioning the transducer at the target with the sCT. The corresponding rCT replaced the sCT skull and the output files were exported. For each virtual targeting procedure (N=20), we calculated the SDR, the ST (the distance between skull boundaries along a ray path), and the NAE between the rCT and sCT. The SDR and ST were averaged across all active elements returned from Kranion. Similarity and statistical significance was determined with Pearson's correlation coefficient and Wilcoxon signed-rank test ($\alpha$ = 0.05) for metrics derived from rCT and sCT.

\subsubsection{Acoustic Simulation using k-Wave}
Once the transducer was positioned about the respective Vim target, the CT image, MR image, transducer element positions, and focus position from the Kranion scene was imported into MATLAB to run full-wave acoustic simulations with the open-source toolbox k-Wave\cite{treeby2010k}. k-Wave solves the first order acoustic wave equation using the k-space pseudospectral method. The CT and MR images were resampled using the imresize3 function in MATLAB (Mathworks, Natick, MA, USA) to an isotropic grid spacing of 0.52 mm and was further zero padded to a grid size of [Nx,Ny,Nz] = [600, 600, 500] to ensure all transducer elements were inside the simulation space. A threshold of 400 HU was used to extract a skull mask from the CT image and an upper limit of 2000 HU was applied to account for the high radio density contribution from any implants in the skull. The intracranial space was set to a constant value of brain tissue and remainder of the simulation grid that contained the transducer was set as water, where Table \ref{Table1} shows the respective acoustic parameters used for all simulations\cite{aubry2003experimental,duck,constans2017200}. The heterogeneous skull layers were incorporated in simulations using a linear approximation to map HU to bone porosity and related porosity to the speed of sound, density and absorption with the following equations\cite{aubry2003experimental}: 

\begin{equation}  \label{eq:1}
    \Phi = 1 - \frac{HU}{max(HU)}
\end{equation}
\begin{equation} \label{eq:2}
    \rho_{skull} = \rho_{water}\Phi + \rho_{bone}(1-\Phi)
\end{equation}
\begin{equation} \label{eq:3}
    c_{skull} = c_{water}\Phi + c_{bone}(1-\Phi)
\end{equation}
\begin{equation} \label{eq:4}
    \alpha_{skull} = \alpha_{min} + (\alpha_{max}-\alpha_{min})\Phi^{0.5}
\end{equation}

\renewcommand{\arraystretch}{1.2}%
\begin{table}[ht]
\caption{Acoustic properties used for all simulations.}
\centering
\begin{tabular}{ccc} 
\toprule
\begin{tabular}[c]{@{}c@{}}Speed of Sound\\(\emph {$m/s$})\end{tabular} & \begin{tabular}[c]{@{}c@{}}Density\\ \emph{$(kg/m^{3})$}\end{tabular} & \begin{tabular}[c]{@{}c@{}}Absorption\\(\emph {$dB/MHz/cm$})\end{tabular} \\ 
\hline
$c_{water} = 1500$ & $\rho_{water} = 1000$ & $\alpha_{water} =0$\\
$c_{brain} = 1560$ & $\rho_{brain} = 1030$ & $\alpha_{brain} = 0.38$\\
$c_{bone} = 3100$ & $\rho_{bone} = 2200$ & $\alpha_{bone,min} = 0.2$\\
\multicolumn{1}{l}{} & \multicolumn{1}{l}{} & $\alpha_{bone,max} = 8$\\
\bottomrule
\end{tabular}
\label{Table1}
\end{table}

The transducer was modeled using the makeMultiBowl function in k-Wave with an element diameter of 8 mm, radius of curvature of 150 mm, and all elements were directed toward the focus position. Simulations were performed at a frequency of 650 kHz, which maintained a spatial discretization greater than 4.3 points per wavelength in water. Three simulations were performed for each target: a simulation without aberration compensation, simulation with applied time delay derived from skull thickness calculations from Kranion, and a simulation with corrections from time reversal. Simulations without correction were performed with the same input amplitude and phase for all elements. Directly applying phases calculated from Kranion to each element was unsuccessful in restoring the focus compared to the no correction case and similarly observed by Lu et al\cite{histotripsy}. Instead, Lu et al. calculated time delays using the skull thickness calculated from Kranion:

\begin{equation}
    t_{Kranion,k} = (R-d_{skull,k})/c_{water} + d_{skull,k}/\overline{c}_{skull}
    \label{eq:5}
\end{equation}

\begin{equation} \label{eq:6}
    \Delta{t_{Kranion,k}} = t_{Kranion,k} - min_{k}(t_{Kranion,k})
\end{equation}

where $R$ was the radius of the transducer array, $d_{skull,k}$ was the skull thickness from Kranion for each element $k$, $c_{water}$ was 1500 $m/s$, and $\overline{c}_{skull}$ was the mean calculated from each skull after conversion by equations \ref{eq:1} and \ref{eq:2}. Phase correction from time-reversal was performed by placement of a virtual point source at the target location which recorded the time-varying pressure for each element. The respective amplitude and phases were extracted with the extractAmpPhase function (taking an FFT of the signal close to the source signal) in k-Wave and the average phase was calculated for all points of a bowl and subtracted from the initial phase. Only the newly acquired phases were applied without changing the input amplitude. To minimize simulation time, a 100 cycle waveform was used, which was the minimum number of cycles to propagate to the target and return to the transducer. All simulations were run on a Quadro P6000 GPU (NVIDIA Corporation, Santa Clara, CA). The root-mean-squared (RMS) pressure was recorded for each voxel location of the grid. To assess simulation similarity between rCT and sCT, we compared the peak pressure (maximum intracranial pressure) and the pressure at the target. We also characterized the beam properties by evaluating focal shift between the peak the target locations, the distance vector between rCT and sCT, and the focal size and volume. Statistical analysis of simulated pressure fields between rCT and sCT were performed with Wilcoxon signed-rank test ($\alpha$ = 0.05). 

\begin{figure}[t]
\centering{}
\includegraphics[width=0.75\columnwidth]{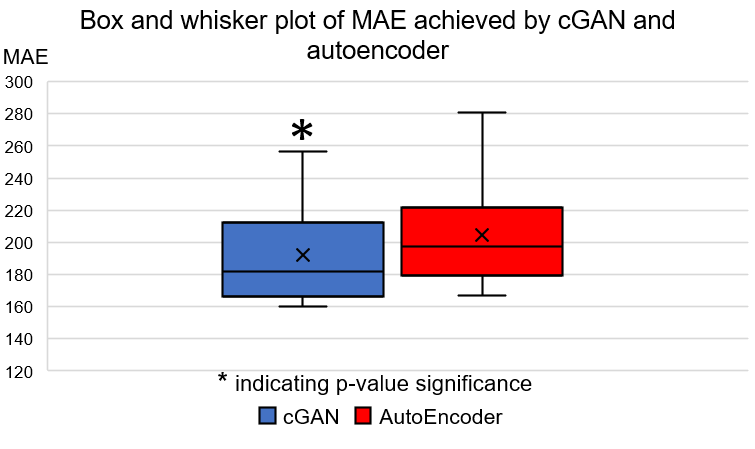}
\caption{\textbf{The box and whisker plot of mean absolute error (MAE) values achieved by cGAN and autoencoder on the testing set.}}
\label{fig1}
\end{figure} 

\begin{figure}[t]
\includegraphics[width=1\columnwidth]{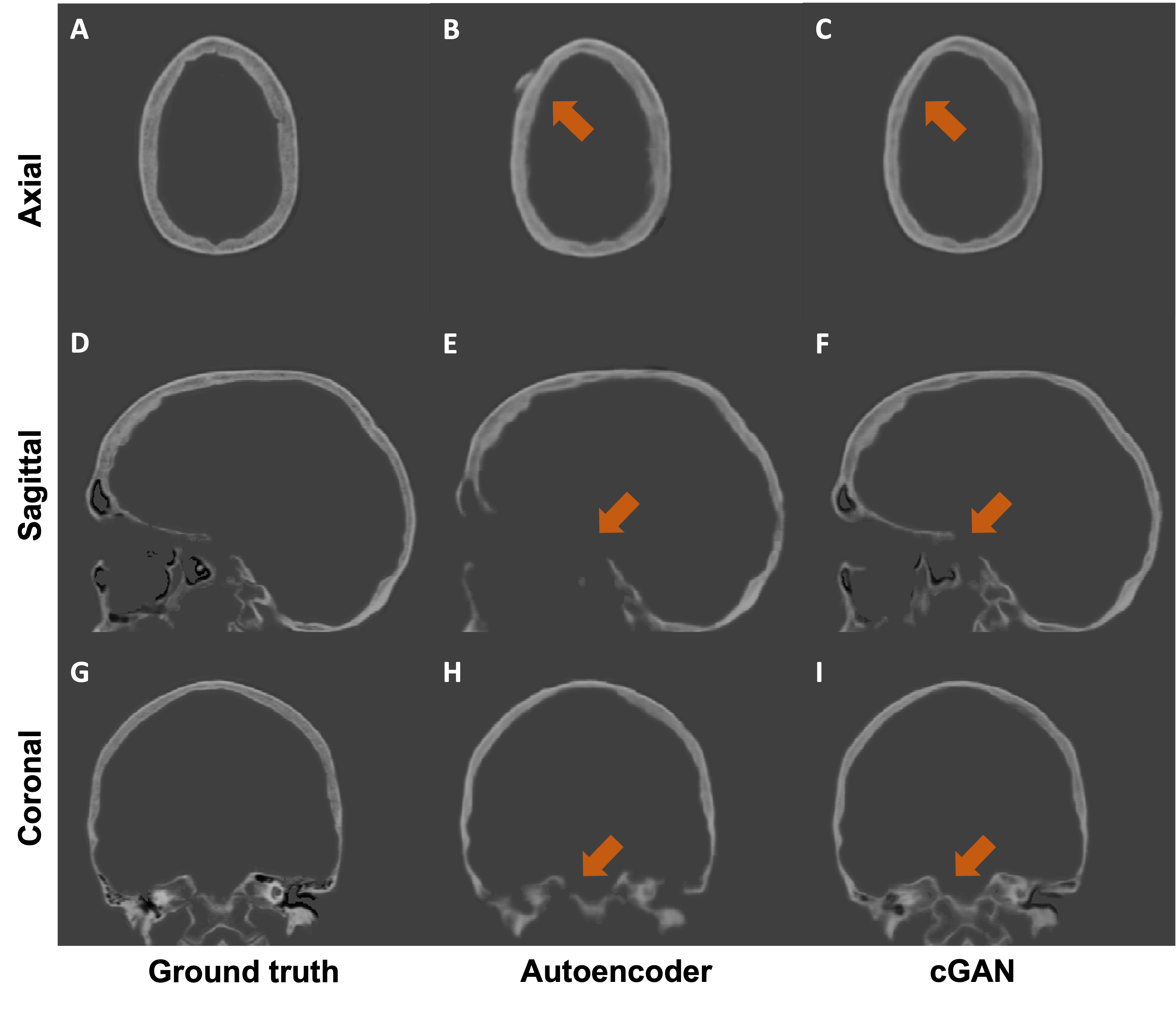}
\caption{\textbf{Qualitative results of an example case from the testing set.} We compare the two mainstream MR-CT translation methods: autoencoder (middle column) and our cGAN (right column). The major differences between two approaches are highlighted by orange arrows.}
\label{fig2}
\end{figure} 

\section{RESULTS}
\subsection{Image Similarity Results}
\begin{figure}[t]
\includegraphics[width=1\columnwidth]{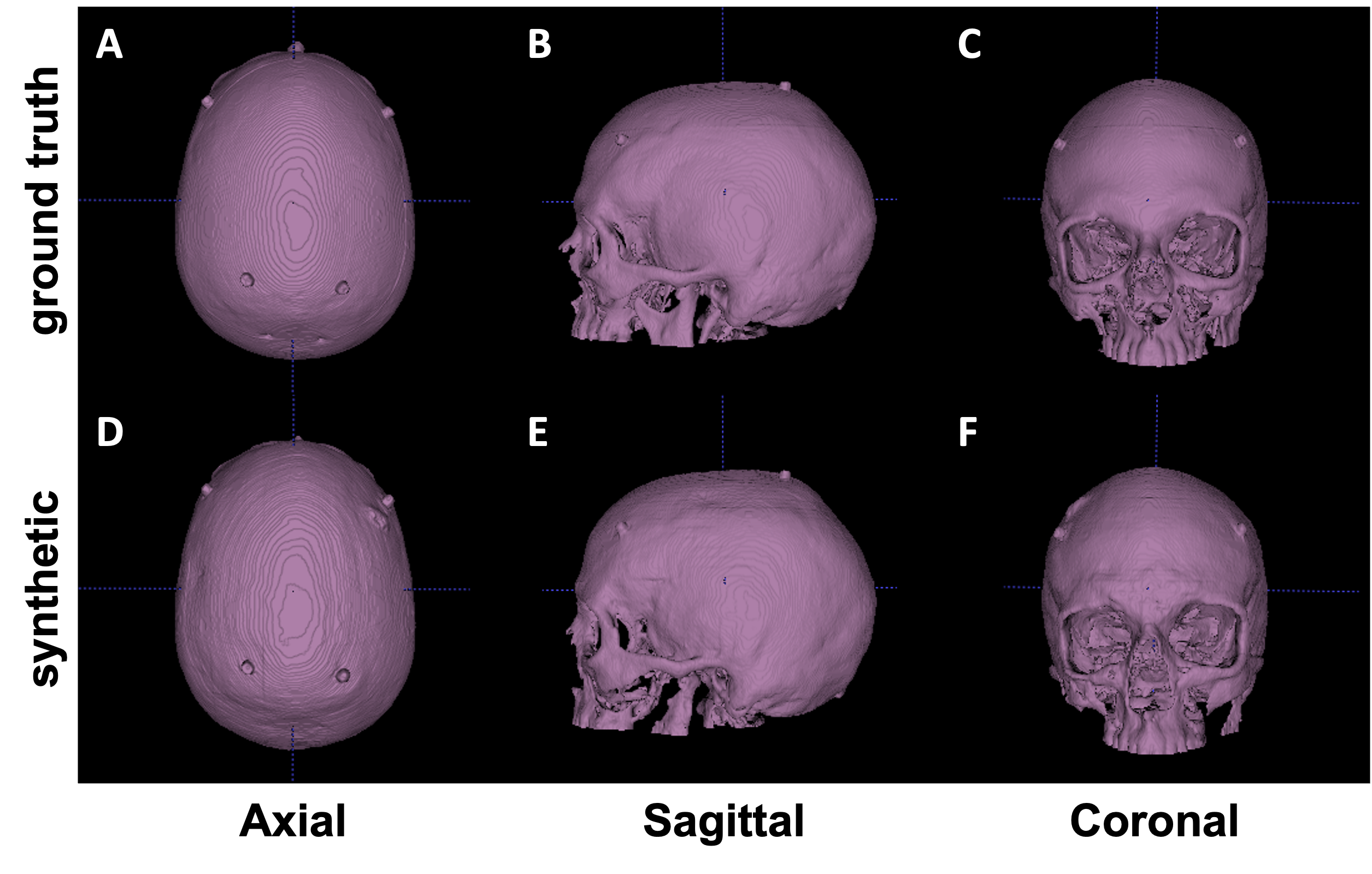}
\caption{\textbf{Visual comparison between real (upper row) and synthetic (lower row) skull.} Isosurfaces emphasize that synthetic CTs are visually comparable to real CTs and contiguous.}
\label{fig3}
\end{figure}

We performed both quantitative and qualitative evaluations of image similarity on our testing set. Quantitatively, we showed the box and whisker plot in Fig \ref{fig1}. Specifically, the MAEs between rCTs and sCTs in skull regions were $192.31\pm28.21$ HU and $206.83\pm27.91$ for cGAN and autoencoder, respectively. We performed paired t-test and found that the difference in mean MAE was statistically significant (p-value $<0.01$). Qualitatively, as shown in Fig \ref{fig2}, we found that at the inferior part of the skull, the synthesized skull exhibited larger difference than at the superior part. The synthetic skull generated by cGAN also included more details and had sharper appearances compared to the one generated by autoencoder. Lastly, we note that our synthesized skulls were spatially continuous in all views and highly comparable to real skulls, as shown in Fig \ref{fig3}.

\begin{figure}[ht]
\centering
\includegraphics[width=1\columnwidth]{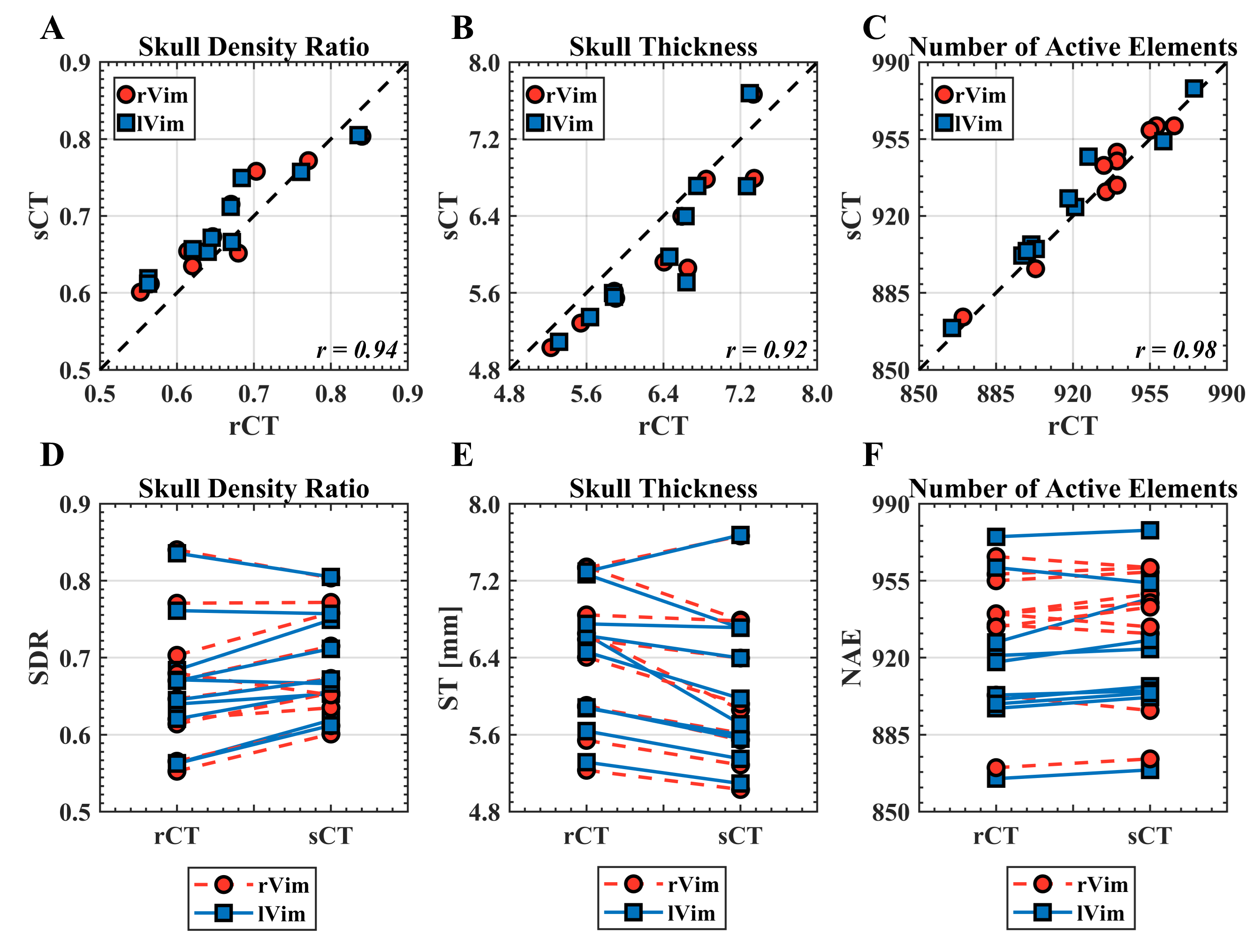}
\caption{\textbf{Kranion-derived skull metrics}. There was a strong linear relationship between the skull density ratio, skull thickness, and number of active elements between the rCT and sCI don't T for both targets (A-C). High Pearson's correlation coefficients shown in the bottom right hand corner of each plot were observed between all three skull metrics. Subplots D-F shows a comparison between rCT and sCT for individual test points, observing that the sCT generally had a higher SDR and a lower skull thickness compared to the corresponding rCT.}
\label{fig4}
\end{figure} 

\begin{figure}[ht]
\includegraphics[width=1\columnwidth]{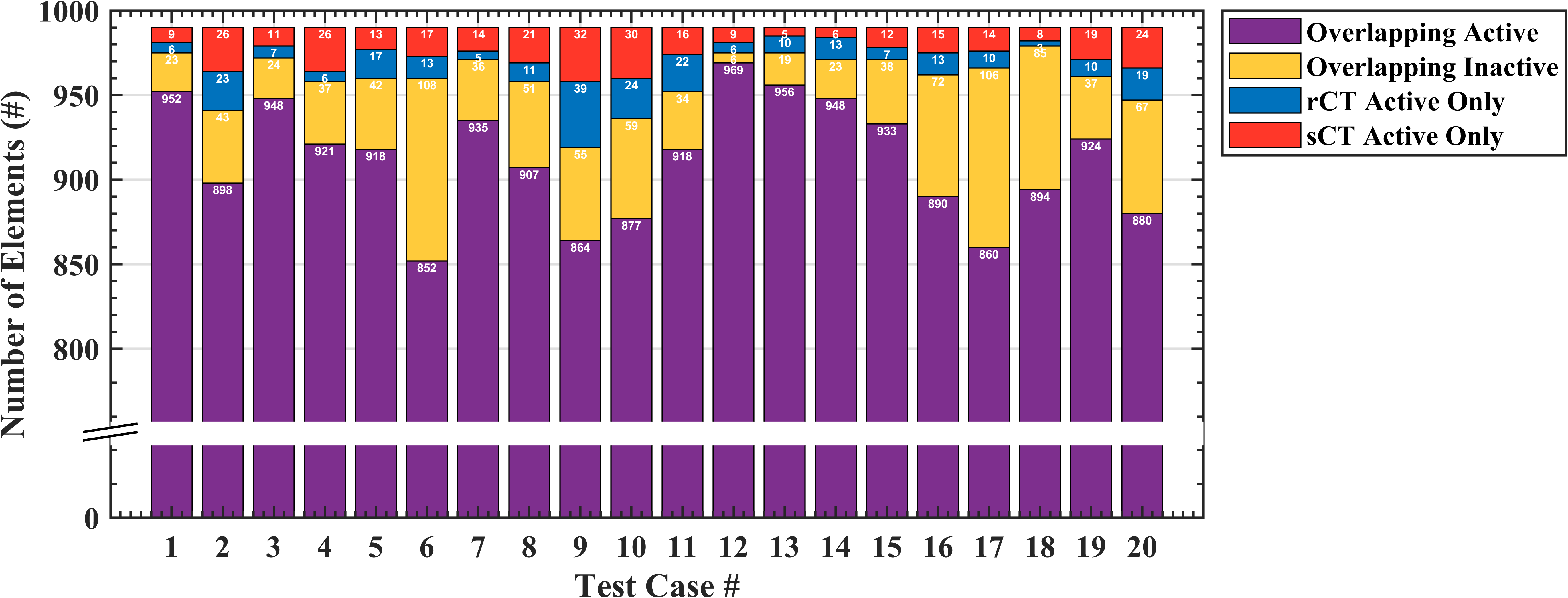}
\centering
\caption{\textbf{Number of active elements.}For all 20 test cases, the active elements calculated from Kranion are compared with the rCT and sCT skulls. Of the 990 elements, the distribution of overlapping active, overlapping inactive, and non-overlapping rCT and sCT active elements are distinguished.}
\label{fig5}
\end{figure} 

\begin{figure}[ht]
\includegraphics[width=1\columnwidth]{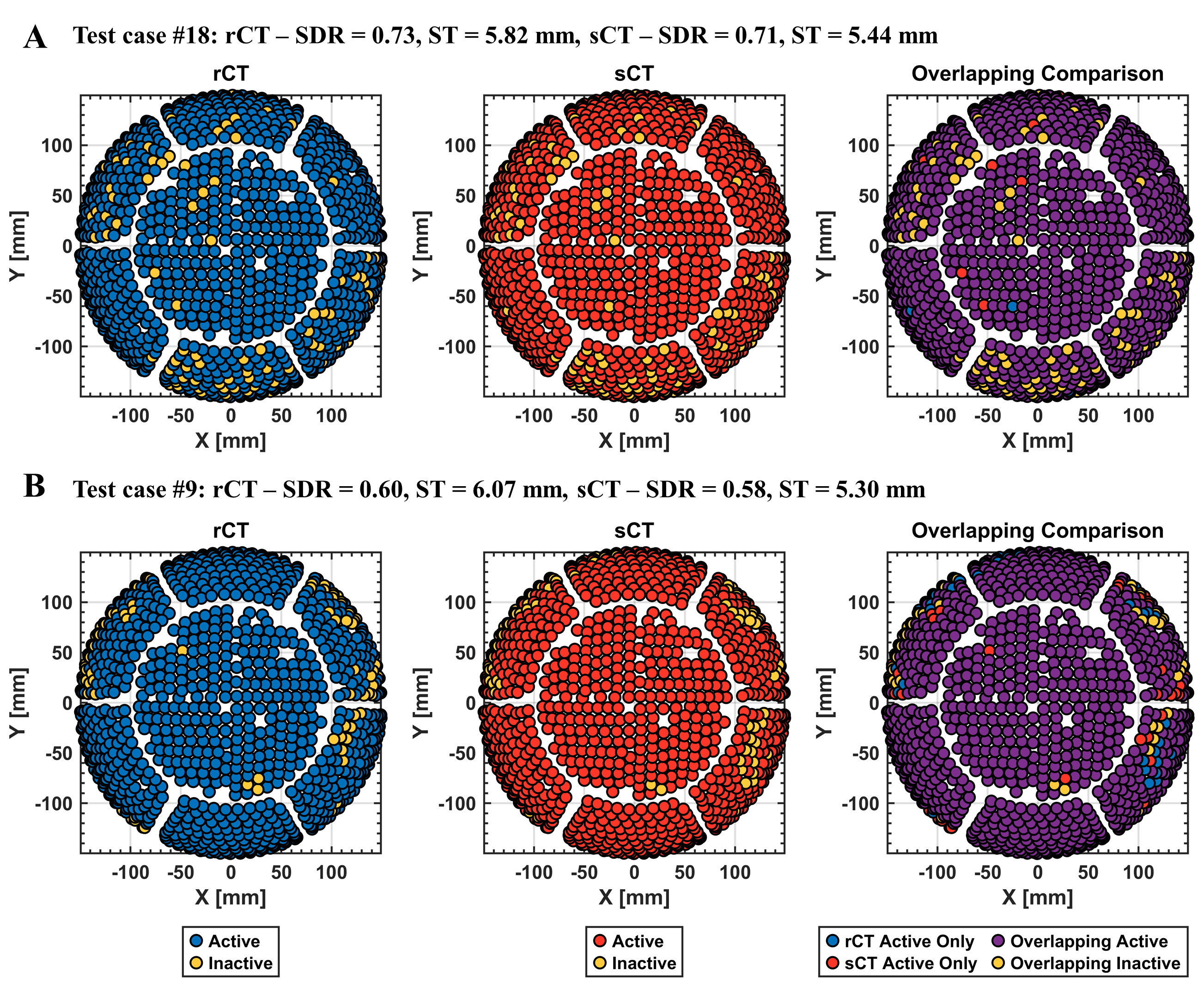}
\centering
\caption{\textbf{Active elements were similar between rCT and sCT.} Visual comparisons are shown for two representative cases to evaluate overlapping active and inactive elements between rCT and sCT. Plots are color coded, showing the distribution of active and inactive elements of the 990 element hemispherical array for the most overlapping case where $98.9\%$ of the active and inactive elements overlap between rCT and sCT (a) and least overlapping case, $92.8\%$ (b). From left to right, elements for the rCT skull, sCT skull, and then compared between rCT and sCT for the overlapping and non-overlapping elements. The case numbers noted in the titles of each subplot correspond with the bar plot in Fig \ref{fig5}.}
\label{fig6}
\end{figure} 

\begin{figure}[ht]
\includegraphics[width=1\columnwidth]{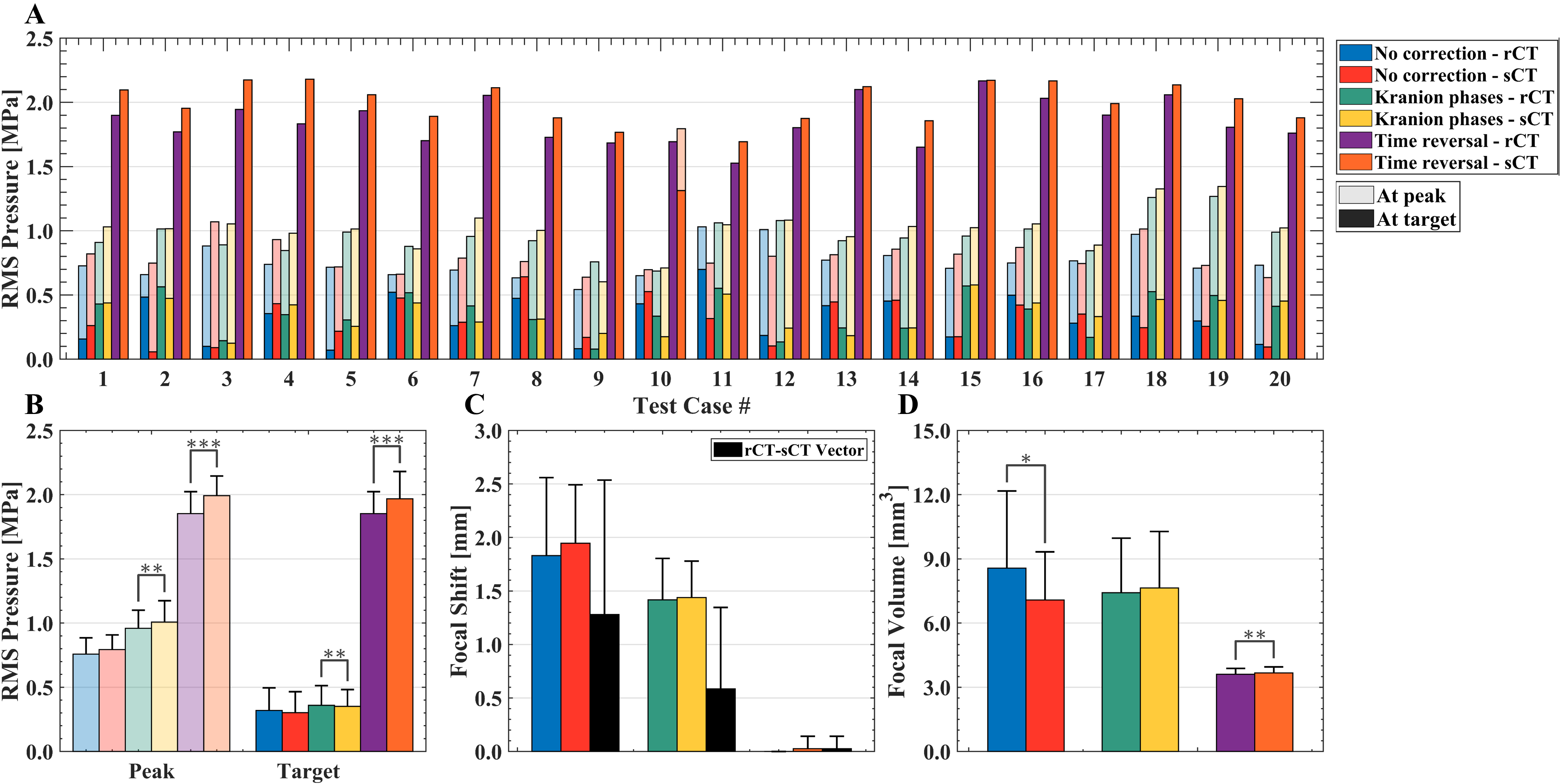}
\centering
\caption{\textbf{Acoustic simulation results from k-Wave.} For each test case, the peak intracranial RMS pressure (transparent bars) and the pressure at the target (solid bars) are shown for rCT and sCT, grouped by phase correction type (A). A group summary of all 20 test cases are shown for RMS pressure, focal shift, and focal volume (B-D).}
\label{fig7}
\end{figure} 

\begin{figure}[ht]
\includegraphics[width=0.9\columnwidth]{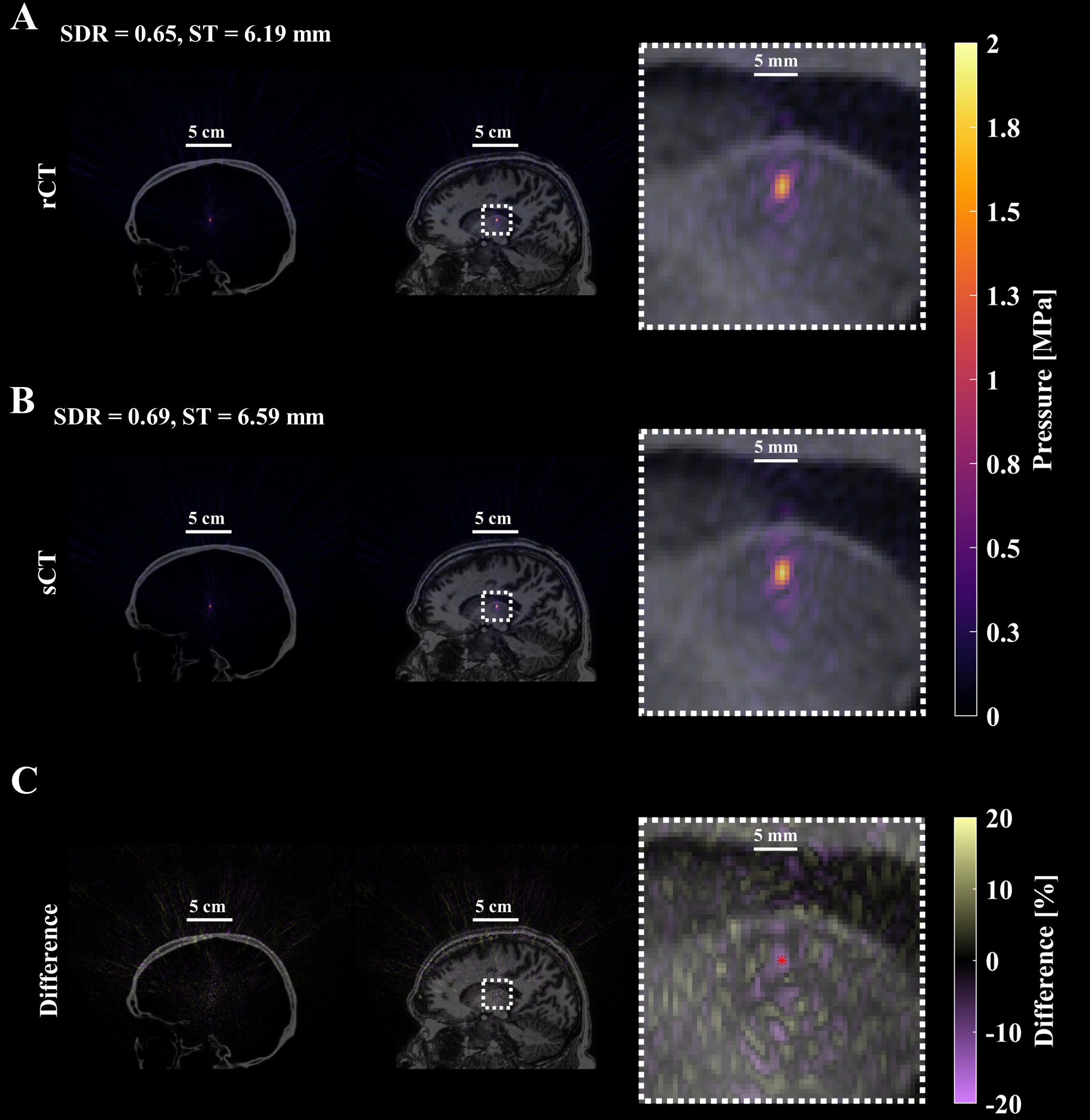}
\centering
\caption{\textbf{Simulated pressure field after applying phases calculated from time-reversal}. An example of the resultant pressure field from one test case with TR-corrected phases targeting the left Vim is shown (test case \#13). The first row contains the pressure field simulated with the rCT skull volume, overlaid on the CT and MR images and enlarged in the final column to better see the focal shape and size (A). Similarly, the sCT skull and simulated acoustic field is presented with similar overlays (B). The percent difference of the pressure fields between rCT and sCT was calculated using the rCT peak pressure as the ground truth and presented in the final row, denoting the intended location with the red dot (C). Qualitatively and quantitatively the spatial extent of the foci are very similar between the rCT and sCT results, with the main difference identified as the maximum pressure.}
\label{fig8}
\end{figure} 

\subsection{Skull Metric Results}
Metrics from Kranion exhibit strong similarity between rCT and sCT across all 20 test cases (Figure \ref{fig4}A-C). The Pearson’s Correlation Coefficients for the skull density ratio, skull thickness, and number of active elements are 0.94 (p $<$ 0.001), 0.92 (p $<$ 0.001), and 0.98 (p $<$ 0.001) respectively, demonstrating strong positive linear correlation of these metrics and significant correlation between rCT and sCT. P-values calculated from the Wilcoxon signed rank test found difference between rCT and sCT derived SDR (p $=$ 0.007) and ST (p $=$ 0.003), but found no difference in NAE between the populations (p $=$ 0.104). Trend lines from paired comparisons between rCT and sCT revealed generally higher SDR and lower ST for sCT compared to rCT. Table \ref{table2} summarizes the mean and standard deviation for SDR, ST, and NAE for the full group comparison. The mean differences between rCT and sCT were $4.8\pm3.7\%$, $5.6\pm3.1\%$, $0.65\pm0.41\%$ for SDR, ST, and NAE. Of the 990 elements that comprised the transducer array, overlapping and non-overlapping elements were compared for each target between rCT and sCT (Fig. \ref{fig5}). On average, $97.0\pm1.5\%$ elements overlapped between rCT and sCT, with $98.9\%$ in the most overlapping test case and the least overlapping test case with $92.8\%$ (Fig. \ref{fig6}A,B).

\renewcommand{\arraystretch}{2}
\begin{table}[b]
\caption{Mean $\pm$ standard deviation calculated for each skull metric from Kranion.}
\centering
\begin{tabular}{cccc} 
\toprule
CT & SDR & ST & NAE\\ 
\hline
rCT & \begin{tabular}[c]{@{}c@{}}$0.65\pm0.07$\\\end{tabular} & $6.4\pm0.66$ & $925\pm30$\\
sCT & $0.67\pm0.06$ & $6.1\pm0.76$ & $929\pm29$\\
\bottomrule
\end{tabular}
\label{table2}
\end{table}

\subsection{Acoustic Simulation Results}
Acoustic simulation results from k-Wave are summarized in Table \ref{table3}. The RMS pressure at the intracranial peak and target locations were compared without phase correction and with applied phases calculated from Kranion and time reversal (Fig \ref{fig7}A). The mean difference in peak pressure between rCT and sCT was $12.4\pm8.1\%$, $7.3\pm6.0\%$, and $7.5\pm10.0\%$ for no phase correction, Kranion phases, and TR corrected phases (Fig \ref{fig7}B). From statistical testing, we observed no difference of peak pressure between rCT and sCT simulations for the case without correction (p=0.062), but observed a difference with Kranion phases (p=0.003) and TR corrected phases (p$<$0.001). Similar relationship were noted for target pressure comparisons between rCT and sCT (no correction: p=0.765, Kranion: p=0.627, TR: p=0.002). The largest distance vector between rCT and sCT peak pressure locations was noted in the case of no correction ($1.3\pm1.2$ mm) and also the largest focal volume difference ($25.5\pm20.5\%$) but both metrics were improved when phase correction was applied (Fig \ref{fig7}C-D). Kranion calculated phases reduced the distance vector of the peak location between rCT and sCT to $0.6\pm0.8$ and focal volume to $20.4\pm23.5\%$. For the TR simulations, there was no focal shift observed between rCT or sCT skulls except for one case where a 0.5 mm offset was observed (test case \#10). TR simulations had the smallest difference in focal volume of $2.4\pm1.9\%$. An example comparing rCT and sCT's pressure fields with TR-corrected phases is observed in Fig \ref{fig8}. 

\renewcommand{\arraystretch}{1.2}
\begin{table}[ht]
\caption{Mean differences between rCT and sCT for metrics calculated from acoustic simulations presented as mean $\pm$ standard deviation. TR: Time-reversal}
\centering
\begin{tabular}{cccc} 
\toprule
                  Simulation & ~Peak Pressure [\%] & Focal Position [mm] & \multicolumn{1}{l}{Focal Volume [\%]}  \\ 
\hline
No correction     & $12.4\pm8.1$        & $1.3\pm1.2$         & $25.5\pm20.5$                                                        \\
Kranion & $7.3\pm6.0$         & $0.6\pm0.7$         & $20.4\pm23.5$                                                                  \\
TR      & $7.5\pm10.0$        & $0.0\pm0.1$         & $2.4\pm1.9$                                                                  \\
\bottomrule 
\end{tabular}
\label{table3}
\end{table}

\section{DISCUSSION and CONCLUSIONS}
Transcranial focused ultrasound is being explored for a number of applications\cite{krishna2018review}. Patient-specific information is required to model effects of the skull during tFUS, and the current gold standard uses a CT to generate a subject-specific map of acoustic properties of the skull. We explored the feasibility of replacing real CT images with synthetic CT images for transcranial focused ultrasound procedures. We hypothesized that results from acoustic simulations with sCT skulls would yield equivalent focal size, location, and pressure compared to simulations with ground truth skulls and evaluated this by comparing the pressure field from three simulation scenarios varying skull compensation methods. We compared skull-derived metrics used clinically to determine patient eligibility using rCT and sCT.  Successful replacement of real CT images with an image generated from routine MR scans could improve tFUS procedure planning by incorporating a subject's skull without causing additional burden on the patient. Through our work here, we showed that the synthesized CT skulls generated by our proposed 3D patch-based cGAN are (1) spatially continuous and (2) highly comparable to real CTs when used to predict skull properties for transcranial ultrasound procedures. 

Image similarity between synthesized and real skulls were first evaluated by qualitative inspection and quantitative assessment. We have explored the effectiveness of two mainstream MR-CT synthesis approaches and demonstrated the superiority of cGAN. Through open-source software, Kranion, we compared clinically relevant skull metrics and active elements derived from rCT and sCT and found they were highly correlated. Statistical testing revealed a difference in skull density ratio and skull thickness, where SDR was slightly overestimated for sCT, while conversely ST was underestimated when compared to rCT. Koh et al. reported correlation coefficients of 0.95 and 0.90 for SDR and ST, comparable to our reported results in this study of 0.94 and 0.92. Acoustic simulations using k-Wave provided a more detailed assessment of ultrasound interaction with the synthetic skulls. Without any skull compensation techniques applied, the peak intracranial pressure difference was higher than that reported of Koh et al. ($12.4\pm 8.1\%$ vs. $3.11\pm 2.79\%$). We speculate this is due to key differences in the simulation setups that include: (1) our simulations were performed at a higher frequency (650 kHz vs. 200 kHz) (2) we modeled the large hemispherical ExAblate array transducer where Koh et al. modeled a single-element transducer. A higher fundamental frequency was used for all transducer modeling in this study to match the clinical system it was modeled after. Because attenuation is frequency dependent, at higher frequencies we expect a greater decrease in intensity at the focus\cite{pichardo2010multi}. The hemispherical array has a much larger surface area than the smaller, single-element transducer (radius of curvature of 150 mm vs ~55 mm), thus a greater amount of skull area influences the simulation outcome, which increases sensitivity to skull differences. 

Our study simulated 3 different tFUS scenarios: no correction for phases offset by the skull, corrections calculated from ray-tracing, and modeling of a virtual source followed by time-reversed phases to restore pressure at the target. TR corrected simulations had the smallest mean difference focal position and volume between rCT and sCT, but application of Kranion-calculated phases had the smallest peak pressure difference. Because the Kranion phases were dependent on the skull thicknesses from Kranion, which we found to be highly correlated between rCT and sCT, we expected the simulation results from applying the phases to be similar. The mean phase difference was smaller for phases calculated from Kranion than TR-corrected simulations, but the TR-corrected simulations more fully encapsulate the internal structure of the skull\cite{aubry2003experimental}. Similar work has evaluated the skull correction performance capabilities of artificial skulls and applied the phases in an experimental context beyond modeling and simulations\cite{wintermark2014t1,miller2015ultrashort,leung2022comparison}. One study reconstructed a virtual CT from a T1-weighted MR image and compared calculated phases from Insightec's ExAblate system, resulting in an average phase difference between the real and MR-generated CT of less than 1 radians and application of the phases was successfully demonstrated in thermal experiments using head phantoms\cite{wintermark2014t1}. A study using the same transducer as Wintermark et al. compared UTE-derived MR images to real CT images of head phantoms and found no statistical difference between peak temperatures achieved for thermal experiments\cite{miller2015ultrashort}. Most recently, Leung et al. generated skulls from UTE images and calculated phases from the Insightec system but applied phases in a water bath setup, measuring the pressure field behind a skull using a hydrophone and reported comparable beam profile results when compared with real CTs\cite{leung2022comparison}. 

Regarding peak intracranial pressure differences observed between rCT and sCT, the mean difference computed in our study falls within an expected range of variability of 10\% that was observed through an intercomparison study across 11 simulation tools\cite{benchmark}. Because the modeled array was large, the grid size necessary to fit all elements would be [Nx,Ny,Nz] = [960,960,540] to maintain greater than 6 PPW in water at 650 kHz and satisfy 3D convergence testing requirements to avoid simulation instabilities.\cite{accsims}. For a single time-reversed simulation, the total run-time on a CPU is approximately 26.9 hours. We instead opted to use GPU-accelerated simulations to reduce computation time but due to the GPU's memory constraints, required us to decrease the spatial resolution to 4.3 PPW. Although the coarser spatial resolution was used for all results reported in this work, we ran a single test case at 0.33 isotropic voxel size (approximately 7.4 PPW in water) to quantify simulation differences without phase correction and with TR-corrected phases for rCT and sCT (N=4). All low-resolution simulations underestimated pressure when compared to high-resolution simulations shown in Supplementary Materials (Figure S1), where the average difference between high and low resolution peak intracranial pressure was $42.4\pm2.5\%$. We acknowledge this pressure discrepancy is high, but we note a relative peak pressure difference between rCT and sCT was similar (high resolution = $6.6\pm5.1\%$, low resolution = $6.5\pm0.5\%$ (raw values shown in Table S1)). Similar focal volume differences between rCT and sCT was observed for high and low resolution simulations shown in Table S2. While the distance vector between rCT and sCT was larger when calculated with high resolution simulation results for the no correction case, we think this transducer's focus is highly aberrated without phase corrections applied. With TR-correction applied we found the vector was offset by a single spatial step size for high and low resolution simulations. Overall, CPU simulations resulted in an underestimation of pressure, but most measurements in the present study do not incorporate these pressure estimates. 

The performance of cGAN generate skulls may be improved by incorporating additional MRI contrast like ZTE into the training process. Recent work evaluated CTs generated from learned T1-weighted MR images, zero-echo time MR images, and direct conversion from ZTE to HU\cite{classical}. Tested through acoustic analysis, ZTE outperformed learned T1-weighted and direct ZTE conversion images for four regions in the brain with low variation. This work suggests ZTE or other imaging sequences can be integrated with convolutional neural networks and tested to observe improvements in similarity between skulls.

Our work evaluated CT images synthetically generated from 3D cGAN with real CT scans for potential use in tFUS thermal and nonthermal applications for a multi-element array. Our study found sCTs generated from the 3D cGAN network could replace the need for CT scans with a routine T1-weighted MR image. Patient selection for tFUS is assessed by metrics characterizing the skull, which we showed sCTs are comparable to rCTs for all skulls. Good similarity was demonstrated for three acoustic simulation scenarios that may arise for tFUS applications. Replacement of rCTs with sCTs would decrease patient burden of additional scan time, minimize exposure to radiation, and eliminate image registration errors.

\section{ACKNOWLEDGEMENTS}
This work has been supported by NIH U18EB029351 and the Advanced Computing Center for Research and Education (ACCRE) of Vanderbilt University. All acoustic simulations were run on a Quadro P6000 GPU donated by NVIDIA Corporation. The content is solely the responsibility of the authors and does not necessarily represent the official views of these institutes. 


\bibliography{references}   
\bibliographystyle{spiejour}   


\vspace{2ex}\noindent\textbf{Han Liu} received his B.S. degree in biomedical engineering and electrical engineering at Rensselaer Polytechnic Institute. He received his M.S. degree in biomedical engineering at Yale University. He is currently pursuing Ph.D. degree in Computer Science at Vanderbilt University. His research interests are in the broad areas of computer vision, deep learning, and medical image analysis.

\vspace{2ex}\noindent\textbf{Michelle K. Sigona} received a B.S. degree in biomedical engineering from Arizona State University in 2017. She is currently pursuing her Ph.D. degree in Biomedical Engineering at Vanderbilt University. Her research interests include transcranial focused ultrasound therapies and acoustic simulations of propagation through the skull. 

\vspace{2ex}\noindent\textbf{Thomas J. Manuel} is a Ph.D. student developing therapeutic ultrasound for neuromodulation and drug delivery to the brain. His interest extend also to cavitation monitoring systems, MRI guidance of transcranial ultrasound procedures, and simulations of transcranial ultrasound scenarios.

\vspace{2ex}\noindent\textbf{Li Min Chen}, M.D., Ph.D., is Professor of Radiology and Radiological Science. Her research interest is to use neuroimaging, neuromodulation, transcranial electrophysiology, and tracer histology to dissect touch and pain circuits in the brain and spinal cord in animal models. She has served as PI on NIH and DoD grants and on NIH and NSF grant application review panels.

\vspace{2ex}\noindent\textbf{Benoit Dawant}, is a Cornelius Professor of Engineering and a Professor of Electrical and Computer Engineering at Vanderbilt University. He is an IEEE Fellow and the Director of the Vanderbilt Institute for Surgery and Engineering. His areas of expertise are medical image processing and analysis, and image-guided surgical procedures . Over the last 20 years he and his laboratory have developed systems for the automatic segmentation of medical images, techniques for evaluating these algorithms, and they have translated these methods to the clinical world. Recent and current projects include the development of techniques to assist in the planning, implantation, and programming phases of Deep Brain Stimulation (DBS) procedures used to treat movement disorders; the development of techniques to assist in the placement and programming of cochlear implants used to treat profoundly deaf patients; and the development of algorithms for the automatic analysis of skin images.

\vspace{2ex}\noindent\textbf{Charles F. Caskey}, Ph.D., has worked in the field of ultrasound since 2004. He received his doctoral degree for studies about the bioeffects of ultrasound during microbubble-enhanced drug delivery under Dr. Katherine Ferrara at the University of California at Davis in 2008. He currently leads an ultrasound laboratory at the Vanderbilt University Institute of Imaging Science where his group focuses on developing new uses for ultrasound, including neuromodulation, drug delivery, and functional imaging.



\end{spacing}
\end{document}